\documentclass[a4paper,fleqn,usenatbib,useAMS]{mnras}

\usepackage{graphicx}	
\usepackage{amsmath}	
\usepackage{amssymb}	
\usepackage{multicol}        
\usepackage{bm}		
\usepackage{pdflscape}	
\usepackage{color}
\usepackage{ulem}

\newcommand{\msol}{M$_{\odot}$}

\newcommand{\ha}{H$\alpha$}
\newcommand{\hb}{H$\beta$}

\newcommand{\heiiwr}{He\,{\sc ii}~$\lambda~4686$}

\newcommand{\niiiwr}{N\,{\sc iii}~$\lambda\lambda~4634/41$}
\newcommand{\nvwr}{N\,{\sc v}~$\lambda~4603/20$}
\newcommand{\heineb}{He\,{\sc i}~$\lambda~4713$}
\newcommand{\niineb}{[N\,{\sc ii}]~$\lambda~5755$}
\newcommand{\civwrr}{C\,{\sc iv}~$\lambda\lambda~5801/12$}
\newcommand{\civheir}{He\,{\sc i}~$\lambda~5876$}
\newcommand{\ciiiwrb}{C\,{\sc iii}~$\lambda~4647/66$}
\newcommand{\civwrb}{C\,{\sc iv}~$\lambda~4658$}

\newcommand{\ergs}{erg\,s$^{-1}$}

\newcommand{\hei}{He\,{\sc i}}
\newcommand{\heii}{He\,{\sc ii}}
\newcommand{\ciii}{C\,{\sc iii}}
\newcommand{\civ}{C\,{\sc iv}}
\newcommand{\nii}{N\,{\sc ii}}
\newcommand{\niii}{N\,{\sc iii}}

\newcommand{\nv}{N\,{\sc v}}
\newcommand{\ovi}{O\,{\sc vi}}

\newcommand{\kms}{\,km\,s$^{-1}$} 

\usepackage[T1]{fontenc}
\usepackage{ae,aecompl}

\title[Wolf--Rayet stars in M\,81 using GTC/OSIRIS]{Wolf--Rayet stars in M\,81 using GTC/OSIRIS: 7 new detections, analysis and classification of the full sample}
\author[V.\,M.\,A.\,G\'omez-Gonz\'alez et al.]{V.\,M.\,A.\,G\'omez-Gonz\'alez$^{1,2,3}$\thanks{E-mail: v.gomez$@$irya.unam.mx}, Y.\,D.\,Mayya$^{1}$, D.\,Rosa-Gonz\'alez$^{1}$,
\newauthor L.\,H.\,Rodr\'iguez-Merino$^{1}$, J.\,A.\,Toal\'{a}$^{3}$ and C.\,Alvarez$^{2}$\\
$^{1}$Instituto Nacional de Astrof{\'\i}sica, \'Optica y Electr\'onica, Luis Enrique Erro 1, Tonantzintla 72840, Puebla, Mexico\\
$^{2}$Universidad Aut\'onoma de Chiapas, Blvd.\,Belisario Dom\'inguez, Km.\,1081, Ter\'an Tuxtla Guti\'errez 29050, Chiapas, Mexico\\
$^{3}$Instituto de Radioastrononom\'{i}a y Astrof\'{i}sica, UNAM Campus Morelia, Apartado postal 3-72, 58090 Morelia, Michoacan, Mexico\\
}

\pubyear{2019}

\begin{document}
\label{firstpage}
\pagerange{\pageref{firstpage}--\pageref{lastpage}}
\maketitle

\begin{abstract}

We report the detection of 7 new Wolf-Rayet (WR) star locations in M\,81 using 
the Multi-Object Spectrograph of the OSIRIS instrument at Gran Telescopio Canarias. 
These detections are the result of a follow-up of an earlier study using the same
instrumental set-up that resulted in the detection of 14 WR locations. 
We analyse the entire sample of 21 spectra to classify them to one of the known WR 
sub-types using template spectra of WR stars in 
the Large Magellanic Cloud (LMC), with similar metallicity to M\,81.
Taking into consideration the dispersion in the 
strengths of the bumps for a given WR sub-type, we found that 19 of the 21 locations
correspond to individual stars, including all the 7 new detections, of sub-types: 
WNL, WNE, WCE and the transitional WN/C.
None of the detections correspond to WCL or WO types.
 The positions of these stars in the 
red bump vs blue bump luminosity diagram agrees well with 
an evolutionary path according to the Conti scenario.
Based on this, we propose this diagram as a straightforward tool for spectral classification of extragalactic WR sources.
The detection of individual WR stars in M\,81, which is at
a distance of 3.6~Mpc, opens up a new environment for testing the massive star evolutionary models.

\end{abstract}





\section{INTRODUCTION}
\label{sec:intro}

Wolf-Rayet (WR) stars are related to some of the most exotic and
interesting astronomical objects in the Universe, e.g. the most
massive stars, binaries, supernova (SN) explosions, compact objects
(neutron stars and black holes), Gamma Ray Bursts (GRBs) and
gravitational waves.
Theoretical models predict that WR stars will end their lives as
powerful core collapse SNIbc, and recent observational works suggest
that rotating WR stars might be also involved in the production of GRBs
\citep[see][and references therein]{2011Vink}. 
Assuming an average WR lifetime:
$t_\mathrm{WR}\sim0.4$~Myr, \citet{2015Moffat}
uses the equation $T=t_\mathrm{WR}/2N$ to estimate
that a random sample of $N\sim$1000 WR stars is needed in order to
expect a SN explosion in the next 200~yrs ($N\sim10^{5}$ for $T=$2~yrs).
The greater the number of WR stars found, the greater the probability
of detecting one of these explosive events in a reasonable human
life-time. 
However, not a single SN, nor a GRB event, has been directly traced back
to a previously classified WR star ($\sim$150~yrs have passed since their discovery).
Moreover, a detailed characterization of individual WR stars
is mandatory to establish a direct connection with theoretical
predictions in massive star evolution and feedback
\citep[][]{2005Meynet,2015Chen,2017Eldridge}.

Extragalactic searches help to increase the WR sample. The advent of
multiple object spectroscopy (MOS) in large telescopes has facilitated
spectroscopic confirmation of large samples of WR star candidates in
nearby galaxies, e.g. M\,31 \citep[0.76~Mpc; ][]{2012Neugent}, M\,33
\citep[0.84~Mpc; ][]{2011Neugent}, NGC\,300 \citep[2~Mpc;
][]{2003Schild} and NGC\,1313 \citep[4.1~Mpc; ][]{2007Hadfield}.
It is important to note that in galaxies that are further away,
the studies of individual WR stars are scarce.

In \citet{2016Gomez} (hereafter Paper~I), we reported the first results of
an observational program for the detection of WR stars in M\,81.
Our spectroscopic targets 
were blue knots in star forming regions in the spiral arms, selected in
archival {\it Hubble Space Telescope} ({\it HST}) images.
We reported the detection of the first population of WR stars in M\,81,
a grand design spiral galaxy at the relatively close distance of
3.63~Mpc \citep[$m - M = 27.8$;][]{1994Freedman},
using longslit and MOS
modes of OSIRIS spectrograph at the 10.4-m Gran Telescopio Canarias (GTC).
The bump strengths in 12 of our 14 detections were consistent with 
those from individual WR stars, comprising of WNL, WNE, WCE and transitional
WN/C sub-types, as inferred using the templates for the
Large Magellanic Cloud (LMC) WR stars.
With the exception of two, all our WR locations were surrounded by ionized
bubbles. The presence of ionized bubbles around WR stars are expected since
they are hot stars supplying both ionizing photons and high velocity 
winds, in addition to being situated in a gaseous environment,
favouring the creation of ionized bubbles. 
Hence we targeted the central stars of a large sample of
ionized bubbles for a follow-up spectroscopic search for WRs with GTC. 

In this paper, we present the detection of new locations with WR features
and a homogeneous analysis of the full sample, dereddened for dust extinction.
In Paper I, we analysed our spectra using templates from WR stars in
the LMC, described in \cite{2006Crowther} and publicly available in the 
website of P.Crowther.
Recent analysis of chemical abundances in M\,81
suggest a uniform abundance in most parts of its disk
\citep[][]{2016Arellano}.
In particular, at galacto-centric distances of 0.35--0.6\,$R_{\rm 25}$,
where our WR stars are located, direct method gives $12+\log$(O/H)=8.10,
which is closer to that of the LMC than that of our Galaxy.
LMC templates are appropriate to analyse the WR features in M\,81 since it
also has examples of WNL, WNE and WCE subtypes.
We also explore a multi-component Gaussian fitting of the broad WR bumps into individual
ionic transitions, in order to construct a red bump (RB) vs blue bump (BB)
luminosity diagram.

This paper is structured as follows: 
candidate sample of ionized bubbles, spectroscopic observations,
reduction, WR identification, extraction and new sample are presented in \S2;
template fitting of M\,81 spectra using LMC
templates are presented in \S3; diagnostic diagrams
are presented in \S4. Finally, summary
and concluding remarks are presented in \S5.

\section{Observations}
\subsection{Sample of central stars of ionized bubbles}

Motivated by the success of detecting individual WR stars through
GTC/OSIRIS spectroscopic observations in M\,81,
we carried out follow-up observations of a carefully
selected candidate sample with the goal of increasing the number of
individual WR stars.
Our previous 14 detections reported in Paper~I
were serendipitous and were located in slits passing through
{\it HST}-selected blue knots in
star-forming regions. Given that 12 of the 14 detected WR stars were
associated with ionised bubbles, we targeted the central stars of
ionised bubbles for these new observations. Bubble-like structures
are expected around WR stars: massive O-type stars evolve into red
supergiant or luminous blue variable stars and lose mass through dense
and slow winds to become WR stars. The strong UV flux from WR stars
ionize the ejected material, giving rise to
shells usually detected through \ha\ narrow-band images
\citep[e.g.][]{1983Chu,2000Gruendl}.
Furthermore, these shells are expected to be swept away by the fast
winds from the WR stars forming the so-called ring nebulae or WR
nebulae \citep[][]{1996Garcia,2011Toala}. The sizes of bubbles in
coeval stellar cluster can reach several tens of parsecs, such as
those observed around our reported WR detections in M\,81 (see
Fig.~1 in Paper~I).

In order to select a sample of candidates for
spectroscopic follow-up, we carried out a systematic visual search for
ionized bubbles in M\,81 using archival {\it HST} images.
The {\it HST} has observed M\,81 with the Advanced Camera for Surveys (ACS),
with the F435W, F606W and F814W broad-band
filters covering the full optical extent of this galaxy.
In particular, we looked for extended structures that are visible only
in the F606W filter. The H$\alpha$ emission line intercepts this
filter, which allows to trace the ionized structures such as bubbles
at a spatial resolution of 0.1\arcsec\ (1.8~pc). A systematic search
resulted in a catalogue of $\sim$300 bubbles with a median radius of
$\sim$12~pc. Around a third of these bubbles contained a bright blue
star within its boundaries. This star is most likely to be either an
individual O-star or a WR star, and hence we considered it as a
primary candidate for positioning the MOS slitlets.
We note that, since the majority of Galactic WR stars are not necessarily
coincident with star-forming regions \citep[e.g.][]{2015Rosslowe}, we do not anticipate completeness
of the WR population through our spectroscopic study of M\,81.

\subsection{Spectroscopic Observations}

\begin{table}
\begin{center}
\caption{\label{tab:log2015} Details of our GTC/OSIRIS MOS
spectroscopic observations${^\dagger}$ in M\,81.}
\begin{tabular}{ccccc}
\hline
Pointing & Date            & Exposure    & Air  & Seeing \\ 
                     &     & Time (s)    & mass & ($''$) \\
\hline
MOS1 & 2015-03-16T22:11:03 &$3\times923$ &1.33 & 0.8 \\
MOS2 & 2015-03-14T22:33:47 &$3\times923$ &1.32 & 0.9 \\
MOS3 & 2015-03-15T22:32:37 &$3\times923$ &1.32 & 1.0 \\
\hline
\end{tabular}
$^{\dagger}$These observations were part of the proposal GTC7-15AMEX 
(PI:\,G\'omez-Gonz\'alez). The location of the 3 pointings are
shown in Fig.~1. The date column also gives the starting Universal Time 
of the $1^{\rm st}$ of the 3~frames, taken one after another.
\end{center}
\end{table}

\begin{figure}
\begin{centering}
\includegraphics[width=1.0\linewidth]{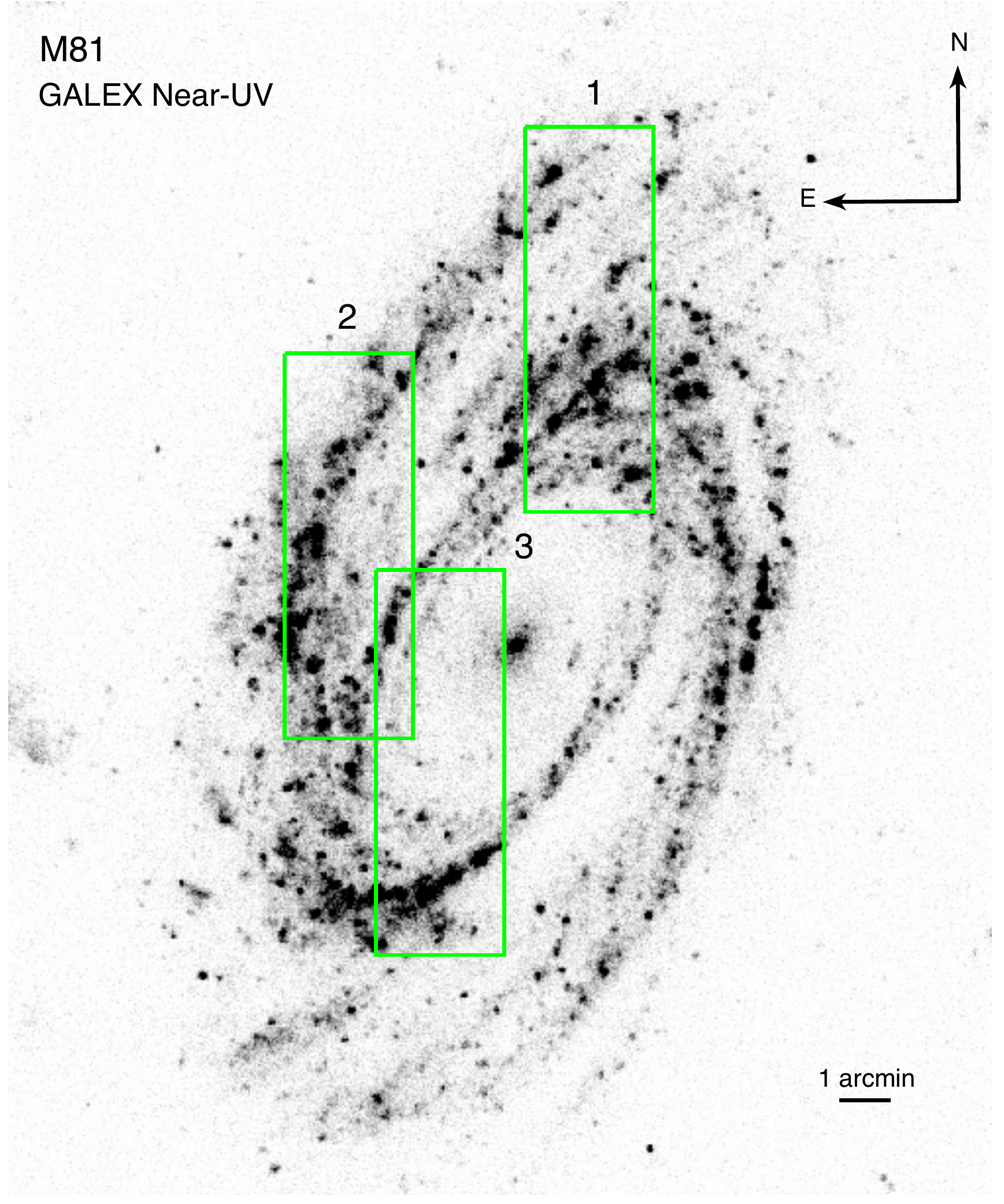}
\caption{\label{fig:galex_image} {\it GALEX} near-UV image of M\,81
showing the location of the 3 GTC/OSIRIS-MOS fields (green rectangles
of $\sim$7.4$^\prime\times2.0^\prime$) analysed in the present work. 
Each FoV contained $\sim$80--90 slitlets of 1.2\arcsec\ width
of varying lengths placed on the central stars of ionized bubbles.
}
\end{centering}
\end{figure}

Spectroscopic observations of the candidate central stars of bubbles were
carried out using the OSIRIS instrument in its MOS mode at the 10.4-m
GTC \citep{2016Cabrera}. Table~\ref{tab:log2015} gives the details of
these observations. Three MOS pointings were used and their locations 
are shown in Figure~1.
Each MOS field of view (FoV) of $\sim$7.4$^\prime\times2.0^\prime$
included $\sim$80--90 slitlets of varying
lengths, including $\sim$5--6 stars as fiducial points
for astrometry. When possible, the slitlets contained object-free
regions for background subtraction. Additionally, we placed around a
dozen slitlets for the purpose of accurate sky/background
determination. Given the geometrical restrictions of mask designer,
we could place slitlets on $\sim$180 central stars of bubbles, our
most likely WR candidates.

We used the same spectroscopic setting as in our earlier observations
presented in Paper~I, i.e., the R1000B grism, slitwidth
of 1.2\arcsec, spectral range of $\sim$3700 to 7500~\AA, spectral
resolution of $\sim$7~\AA\, and a CCD binning of 2$\times$2. The
observations have a spatial scale of 0.254\arcsec\,pixel$^{-1}$ and
spectral sampling of $\sim$2\,\AA\,pixel$^{-1}$.
The observations were carried out in service mode, with the total
observing time split into blocks of $\sim$60~min, each corresponding
to a particular MOS pointing, and consisted of 3 exposures of equal
integration time to facilitate later removal of cosmic rays.
Data from each block contained auxiliary files that included standard
stars, bias, flat-field and arc lamps. The sky was stable during all
the observations with a seeing between 0.8--1.0\arcsec.

\subsection{Reductions, WR identification and extraction}

The procedure adopted for the reduction and extraction of spectroscopic 
data is similar to that described in Paper~I. Reduction was 
carried out using the package {\sc gtcmos}, a tailor-made {\sc iraf}-based 
pipeline\footnote{http://www.inaoep.mx/$\sim$ydm/gtcmos/gtcmos.html} 
developed by Y.~D. Mayya for the reduction of the OSIRIS spectra.
Wavelength and flux-calibrated 2-D spectral images of the 3 fields
are visually examined on a display monitor for the identification of 
the blue bump, a characteristic signature of a WR star.
Out of the 179 spectra of central stars, we identified
7 spectra containing WR features.
The remaining objects are mostly O and early B-stars, which will be the 
subject of study in a future work.

The physical scale of the slitwidth ($\sim$21~pc, at the distance of
M\,81) is large enough to include more than one star. However, our
targets were pre-selected to contain a stellar object in the slit at
the {\it HST} resolution of ($\sim$1.8~pc). This allowed easy
association of a spectra with a detected WR feature to a star on the
{\it HST} image.
In Figure~2, we show colour-composite {\it HST} ACS images
of the 7 new detections.

Bumps of WR\,15, WR\,17, WR\,18 and WR\,20 had an associated
continuum. For each of these WR features, an 1-D spectrum was extracted
using the {\sc IRAF} task {\it apall}, summing spectra over 4--7 pixel,
depending on each case, about the traced continuum.
On the other hand, WR\,16, WR\,19 and WR\,21 did not have an associated
continuum. In these cases, the continuum of a bright object, close
enough to each continuum-less WR, was used as a reference. The {\it reference} 
parameter from task {\it apall} was activated to use the polynomial of
the continuum as reference for extraction.

\begin{table*}
\begin{center}
\caption{\label{tab:newseven}Sample of WR stars in M\,81.}
\begin{tabular}{lcccrrrlrl}
\hline
ID &  R.A. & Dec   & $V$  & $B-V$  & $B-I$ & $M_{V0}$ & Complex &  Size$_{\rm neb}$ & $A_{\rm V}$\\
(1)&    (2)& (3)   & (4)  & (5)    & (6)   & (7)   & (8)     & (9) & (10) \\
\hline
WR\,1   & 9:55:01.663 & +69:12:57.57 & 21.28 &  0.53 &  0.23 & $-6.72 $ &Munch 18 & 100 & 0.2 (Neb) \\
WR\,2   & 9:55:44.299 & +69:07:19.43 & 20.53 &$-0.16$&$-0.25$& $-7.47 $ &R06B06945& 100 & $<$0.5*\\
WR\,3   & 9:55:16.651 & +69:08:55.41 & 19.51 &  0.05 &  0.09 & $-8.49 $ &kauf152  & 250 & 0.5 (Neb) \\
WR\,4   & 9:55:35.890 & +69:07:48.15 & 22.62 &  0.04 &$-1.05$& $-5.38 $ &--       & --  & 1.0 (Neb)\\
WR\,5   & 9:55:09.785 & +69:09:19.71 & 23.11 &$-0.23$&$-0.52$& $-4.89 $ &--       & --  & 1.5 (Neb) \\
WR\,6   & 9:54:42.622 & +69:03:42.67 & 21.56 &$-0.09$&$-0.22$& $-6.44 $ &kauf125  &  20 & 0.2 (Gal) \\
WR\,7   & 9:54:47.525 & +69:04:34.22 & 20.45 &  0.16 &  0.37 & $-7.55 $ &kauf127  &  80 & 0.8 (Neb) \\
WR\,8   & 9:54:46.757 & +69:04:22.51 & 21.43 &  0.05 &  0.11 & $-6.57 $ &--       &  80 & 1.0 (Neb) \\
WR\,9   & 9:54:42.566 & +69:03:29.94 & 22.36 &  0.38 &  0.16 & $-5.64 $ &kauf125  & 100 & 0.2 (Gal) \\
WR\,10  & 9:54:49.418 & +69:07:19.03 & 21.90 &  0.52 &  0.57 & $-6.10 $ &kauf135  &  80 & 1.6 (Neb) \\
WR\,11  & 9:54:48.643 & +69:05:59.65 & 19.28 &  0.16 &$-0.16$& $-8.72 $ &kauf128  &  80 & 0.2 (Gal) \\
WR\,12  & 9:55:22.682 & +69:09:52.26 & 21.79 &  0.35 &  0.20 & $-6.21 $ &--       & 100 & 0.2 (Neb) \\
WR\,13  & 9:55:24.986 & +69:08:14.82 & 20.22 &  0.11 &  0.19 & $-7.78 $ &kauf159  & 250 & 0.8 (Neb) \\
WR\,14  & 9:55:16.488 & +69:08:55.33 & 22.35 &  0.11 &  0.06 & $-5.65 $ &kauf152  & 250 & 0.7 (Neb) \\
WR\,15  & 9:55:28.817 & +69:12:53.97 & 20.46 &  0.09 &  0.18 & $-7.54 $ &Reg55    &  50 & 0.2 (Gal) \\
WR\,16  & 9:56:16.831 & +69:03:36.42 & 23.22 &  0.72 &  1.33 & $-4.78 $ &Reg41    &  20 & 0.7 (Neb) \\
WR\,17  & 9:55:47.042 & +68:59:18.97 & 20.23 &  0.51 &  0.77 & $-7.77 $ &kauf178  &  75 & 1.5 (Neb) \\
WR\,18  & 9:56:06.300 & +68:59:34.35 & 22.56 &  0.13 &  0.24 & $-5.44 $ &kauf232  &  30 & 1.4 (Neb)\\
WR\,19  & 9:56:05.225 & +68:59:55.77 & 23.75 &  0.12 &  0.20 & $-4.25 $ &Reg42    &$<3$ & 0.4 (Neb)\\
WR\,20  & 9:56:00.211 & +69:04:21.60 & 21.48 &$-0.04$&$-0.01$& $-6.52 $ &Reg43    &  50 & 0.2 (Gal)\\
WR\,21  & 9:56:17.969 & +69:05:09.88 & 22.01 &$-0.04$&$-0.18$& $-5.99 $ & --      &$<3$ & $<$0.9*\\
\hline
\end{tabular}\\
\end{center}
Brief explanation of columns:
(1) Name of WR star adopted in this study. Detections WR\,1--WR\,14 are from Paper I,
whereas WR\,15--WR\,21 are new detections;
(2--3) Right Ascension and declination in the FK5 system on the astrometrised 
{\it HST} image, where the M\,81 nucleus is located at R.A.=148.88828$^\circ$ (9:55:33.188) 
and Dec=69.065263$^\circ$ (+69:03:54.95);
(4-6) Apparent magnitude and colours in F435W(B), F606W(V) and F814W(I) bands;
(7) Absolute magnitude in F606W band ($M_{V0}$) using a distance modulus of 27.80~mag,
and Galactic extinction $A_{\rm V}=0.20$~mag \citep{2011Schlafly};
(8) Star-forming complex from \cite{2006Perez} embedding the WR star;
(9) Size in parsec of the nebulosity around the WR star as measured on the F606W image;
(10) Extinction $A_{\rm V}$ (mag). Asterisk(*) indicates an upper limit. See \S2.4 for details.\\
\end{table*}

\begin{figure*}
\begin{centering}
\includegraphics[width=\linewidth]{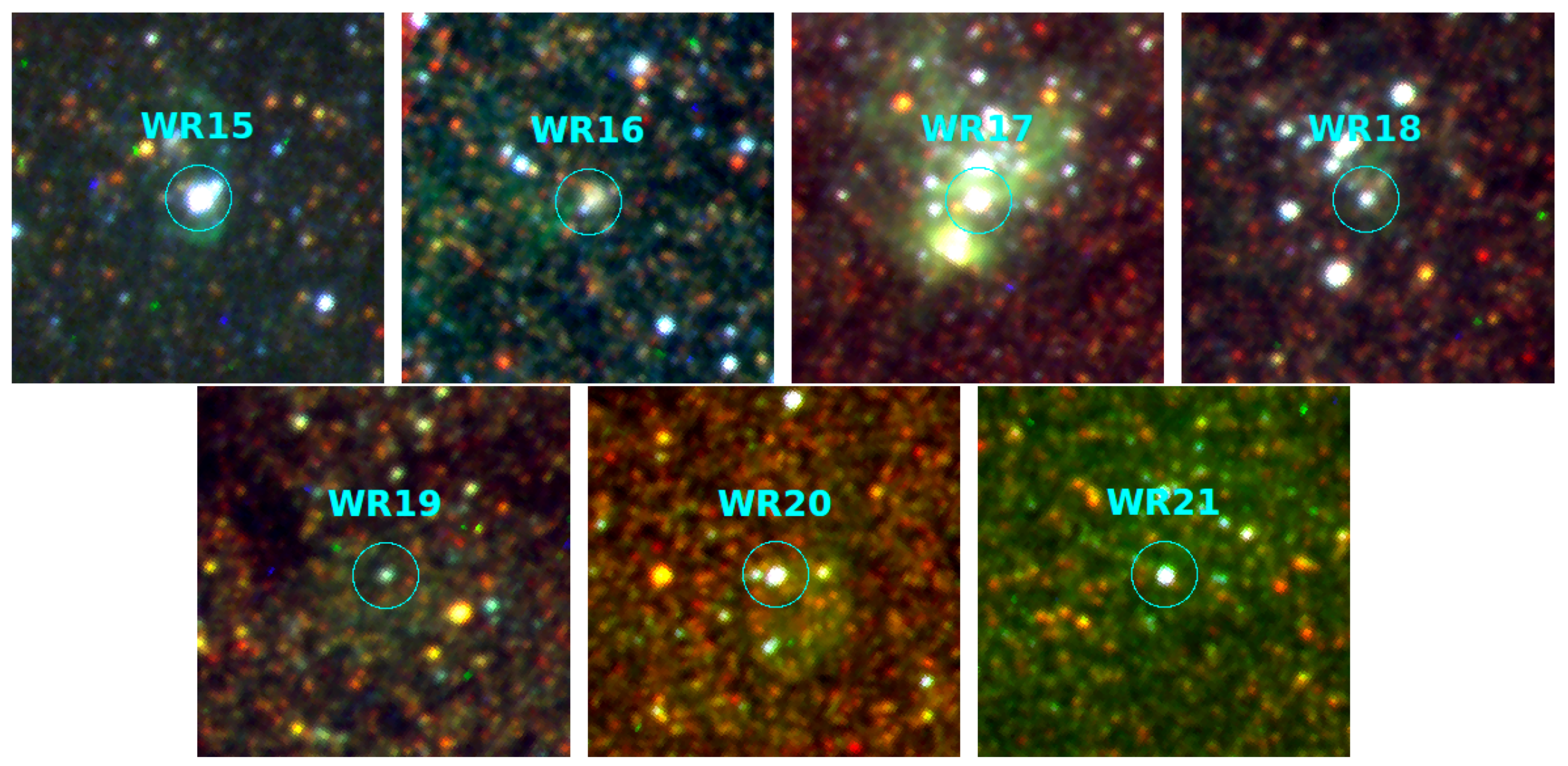}
\caption{100~pc$\times$100~pc colour-composite {\it HST}/ACS 
cut-outs around the 7 new WR locations in M\,81 (WR\,15--21).
Red, green and blue correspond to the {\it HST}/ACS F814W, F606W and F435W filters, respectively. 
The WR stars are identified by circles of 1.0\arcsec\ diameter, which is comparable to the slitwidth
used in these observations. The F606W filter intercepts \ha\ line, which gives a green appearance
to the ionized nebulae and bubbles. North is up and east to the left in all images.}
\end{centering}
\label{fig:fig2}
\end{figure*}

\begin{figure}
\begin{centering}
\includegraphics[width=0.95\linewidth]{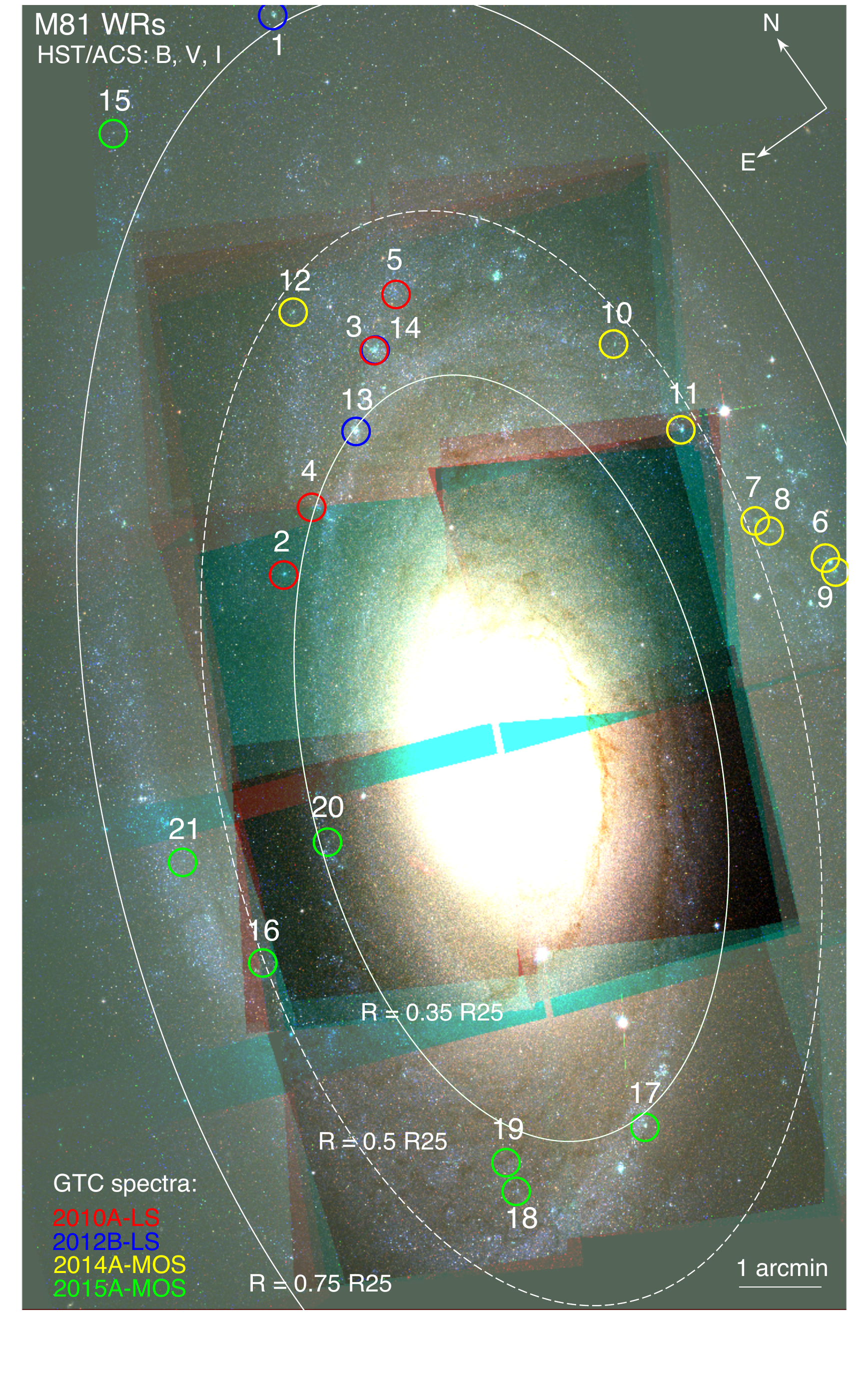}
\caption{Colour-composite mosaic image of M\,81 using {\it HST} 
ACS images in the F814W (red), F606W (green) and F435W (blue) filters
showing the 21 locations (circles) where WR features are detected in our
GTC spectra spanning over 4 observing runs.
Colours red, blue, yellow and green of the circles correspond to 
observing runs 2010A (longslit), 2012B (longslit), 2014A (MOS) and 2015A (MOS),
respectively. The 7 new detections, labelled as WR\,15--WR\,21 in this work,
are indicated with green circles.
Different galactocentric distances are indicated with ellipses of semi-major
axis = 0.35, 0.5 and 0.75 R25.}
\end{centering}
\label{fig:fig4}
\end{figure}

\subsection{The updated sample of WR stars in M\,81}

The 7 new WR detections, labelled as WR\,15--WR\,21, along with 
the previous 14 objects reported in Paper~I, add up to a total of 21 WR
detections in M\,81. Table~\ref{tab:newseven} lists some of their
basic parameters, where the first 14 entries are taken from Paper~I. 
The photometry for the new sources was carried out using
{\sc iraf} task {\it phot} as in Paper~I. We
also list the size of the nebulosity around the WR object, when
present, as measured on the {\it HST} F606W image, and the name of the
star-forming complex, from \cite{2006Perez}, embedding the WR
location.
The locations of these 21 WR sources are indicated in Figure~3, which
is a colour-composite wide-field {\it HST} ACS mosaic
image. As expected, all the WR sources are confined to the spiral
arms, which is hosting many star forming regions.

Individual stars and nebulae in M\,81 are known to suffer a mean extinction of 
$A_{\rm V}\sim0.8$ \citep[e.g.][]{2012Kudritzki,2019Humphreys} which is
much higher than $A_{\rm V}(Gal)=0.20$ along the line of sight to M\,81.
Thus, the observed spectra need to be corrected for reddening. For this purpose,
we determined the visual extinction $A_{\rm V}(nebula)$ using the Balmer decrement method.
We measured the \ha\ and \hb\ emission line fluxes 
in each extracted WR spectrum and obtained $A_{\rm V}(nebula)$ assuming intrinsic
Balmer decrement ratio corresponding to a case B photoionized nebula of 
T$_{\rm e} = 10000$ K and n$_{\rm e} = 100$ cm$^{-3}$
\citep{2006Osterbrock} and the reddening curve of \cite{1989Cardelli}.
The resulting $A_{\rm V}(nebula)$ are listed in the last column of Table~2.
In five cases (WR6, WR9, WR11, WR15 and WR20), $A_{\rm V}$ obtained from
Balmer decrement was of the same order or marginally less than $A_{\rm V}(Gal)=0.20$, 
in which case, we report $A_{\rm V}(Gal)=0.20$ in this column. In two other cases
(WR2 and WR21), $B-V$ colours when dereddened with $E(B-V)=A_{\rm V}(nebula)/3.1$
turned out to be bluer than $-0.32$, the bluest $B-V$ colour of the Planck function,
clearly indicating that the $A_{\rm V}(nebula)$ over-estimates the extinction 
towards these WR stars. For these cases, we give an upper limit calculated as
$A_{\rm V}=3.1\times (B-V + 0.32)$.
Reduced extinction towards these WRs, as compared to the $A_{\rm V}(nebula)$, 
is not unexpected given that the strong winds emanating from the WR stars often create 
gas and dust-free zones in the immediate vicinity of the star. Thus, in general,
$A_{\rm V}(nebula)$ is an upper limit to the extinction of the WR star.
Hence, we dereddened all spectra with two limiting values of $A_{\rm V}$:
(1) maximum represented by $A_{\rm V}(nebula$) and (2) minimum given by $A_{\rm V}(Gal)=0.20$.
Both these dereddened spectra were analysed in \S3.1 to determine its WR content.

\section{Classification of WR stars}

WR stars are identified and classified using the relative strengths
of various transitions of nitrogen (N), carbon (C) and oxygen (O) ions, in addition
to that of doubly ionized helium (\heii). Most of these transitions are located
around two wavelengths, commonly known as the blue bump (BB; $\sim$4600--4700\,\AA)
and the red bump (RB; $\sim$5800--5840\,\AA).
The BB contains one or several \heii, \niii, \nv\ and \civ\ lines,
whereas the RB is made mainly of \civ\ lines.
WR stars exhibiting any N line are classified as WN type,
whereas those containing a C line are of WC type. Rarely a WR 
spectrum of a Galactic star contains both the N and C lines. These 
rare cases have been named as transitional WN/C sub-types \citep{1989Conti}.
Each sub-type is further classified into early (hereafter WNE and WCE) or 
late (hereafter WNL and WCL), depending on whether the detected
features are of high or low ionization species, respectively. The
emission lines of both the high and low ionization states of N are
located in the BB and those of C in the RB, at separable wavelengths.
Thus, the location of the bumps in the spectra allows the
classification into early and late types. The \heii\ line is seen in
both the WN and WC types. Furthermore, in the WC phase, lines are
systematically broader than in the WN type \citep{1989Conti}.
All these well-known characteristics make it possible to classify WR stars
into its sub-types just by the analysis of their BB and RB
\citep{1968Smith, 1996Smith, 1998Crowther}.

Unlike the Galactic WR stars, typical slitwidths used in spectroscopic 
observations intercept surrounding nebulae around extra-galactic WR stars.
In addition, some of the ionic lines are faint to be detected at the
distance of M\,81 even with the 10-m class telescopes. This inhibits a
direct classification into WR subtypes from the observed spectra.
This limitation is overcome by using the template spectra for classification
of the observed extra-galactic WR spectra 
\citep[e.g.][]{2006Hadfield,2007Hadfield,2013Kehrig}.
The most commonly used templates 
for this purpose are from the observations of WR stars in the LMC.
Comparison of the strengths of WR features in the template and observed
spectra gives an indication of whether the
observed feature belongs to an individual star or to multiple stars.
We use the template fitting method to classify our WR spectra in M\,81,
which is described below.

\subsection{Template fitting of M\,81 spectra}

\begin{figure*}
\begin{centering}
\includegraphics[width=\linewidth]{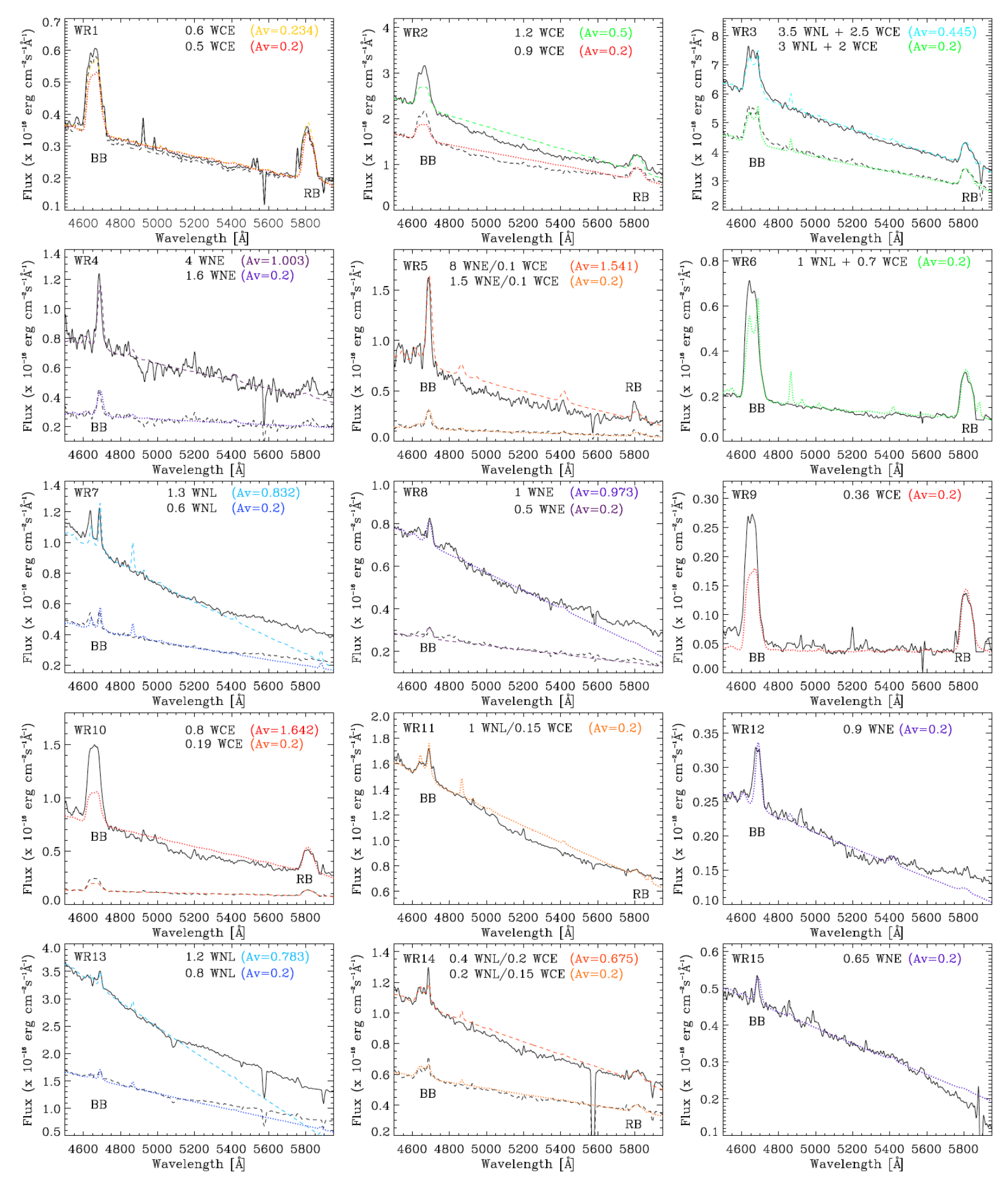}
\par\end{centering}
  \caption{
Observed spectra of 21 WR locations in M\,81 WRs (solid line),
along with their best-fit LMC templates (coloured dotted line).
WR name (WR\,1--WR\,21), name of the template (WNE, WNL, WCE)
and the multiplicative factor required to match the observed bump
strengths are indicated in each panel.
Nebular spectra were subtracted for the
first 15 using the spectra of pixels adjacent to the WRs, whereas
such subtraction was not possible due to
the short slitlengths for the last six spectra (for WR\,16 -- WR\,21).
Note that two values of $A_{\rm V}$:
(1) $A_{\rm V}(nebula)$ and (2) $A_{\rm V}(Gal)=0.20$
are considered in order to correct our spectra for dust extinction.}
\end{figure*}

\begin{figure*}
\begin{centering}
\includegraphics[width=\linewidth]{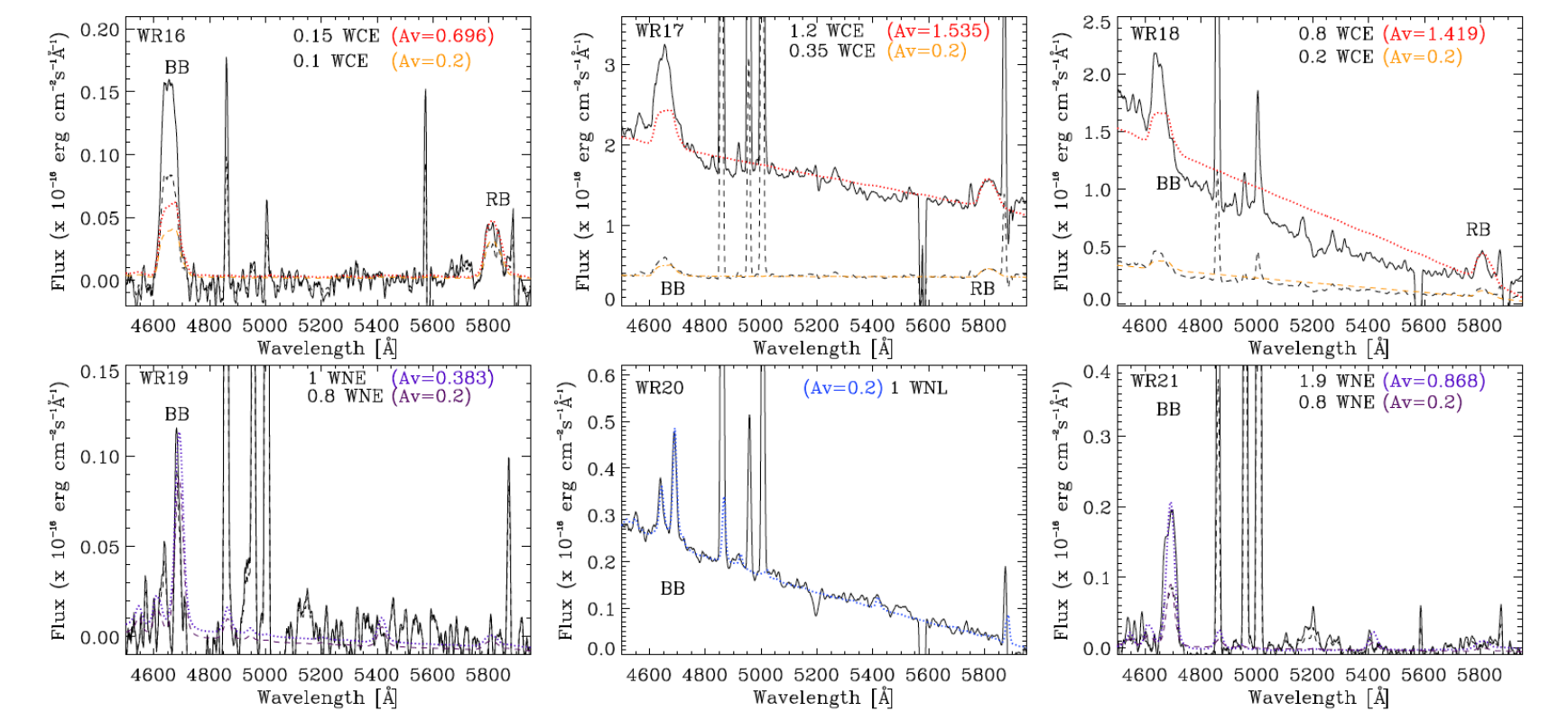}
\par\end{centering}
Fig.~4 continued.
\end{figure*}

Given that the metallicities obtained by direct method in M\,81 \citep[][]{2016Arellano} 
compare well with that of the LMC, we
use the LMC templates presented in \citet{2006Crowther}. These are
available\footnote{http://www.pacrowther.staff.shef.ac.uk/science.html}
for 4 sub-types of N-rich WRs: WN7-9, WN5-6, WN4-6 and WN2-4, 
and one sub-type of C-rich WRs: WC4.
Class numbers greater than or equal to 6 are
usually referred to as late and the rest being called as early.
Thus, we have three WNL, and 1 WCE LMC templates.
The template spectrum for each sub-type is an
average of many stellar spectra of the same sub-types and are 
given in luminosity units. All the WRs in the LMC are at the same
distance and hence the LMC templates are not affected by the errors
in distance determination.

The template spectra were first scaled so that both the observed and
template spectra are in flux units at the distance of M\,81. We
superposed the template spectrum of a given sub-type on an observed
spectrum and visually examined the relative strengths of bumps in the
two spectra. 
For a spectrum to be considered to belong to a candidate WC star,
it should not only have a detectable RB, but also the sum of the 
luminosities of the two bumps should be at least $5\times10^{35}$~\ergs. 
For these cases, we started with a template of WCE, and then applied 
a multiplicative factor on it so that the RB in the observed and 
template spectra are well matched.
We then examined the BB. If the continuum levels at the BB are
different, we added a pseudo-continuum (a straight line with a
variable slope) to the template spectrum so that the observed and
template spectra have the same slope between the bumps. If the
observed BB shows an excess, especially on its bluer edge, it suggests
the presence of a N line. We then found a multiplicative factor
required to fit the BB, either with a WNL or a WNE template. The
procedure is repeated until the observed bumps are well matched by a
single, or a combination of templates. In cases where there is no
detectable RB, or the sum of the bump luminosities is $<5\times10^{35}$~\ergs, 
only the WNL and WNE templates are used.

The results of fitting LMC templates for our sample of 21 WR spectra,
each dereddened using $A_{\rm V}(nebula)$ and $A_{\rm V}(Gal)$
and \citet{1989Cardelli} reddening curve are presented in Figure~4. 
The inferred WR type as well as the multiplicative factor ($f_{\rm WR}$) that had to be applied to 
the template so that it matches the observed bump strengths for both the values 
$A_{\rm V}$ are indicated in each plot. 
As expected, dereddening increases the flux of the BB relative to that of the RB.
Thus, $f_{\rm WR}$ is systematically higher for spectra dereddened 
with $A_{\rm V}(nebula)$ as compared to that with $A_{\rm V}(Gal)$. 
For WNL and WNE stars, the BB profile structure, as well as the observed intensities, 
are well-reproduced by the LMC templates.
On the other hand, observed BB strength of M\,81 WCEs is systematically higher as compared to that
of the LMC WCE template, suggesting that this template is not adequate to the M\,81 WCEs. 

The $f_{\rm WR}$ is directly related to the number of WR stars contributing
to the analysed spectrum. The inferred number is often less than 1 even 
for spectra dereddened with $A_{\rm V}(nebula)$.  
This is expected as not all stars of the same sub-type have exactly the same
bump strengths. In fact, there is almost an order of magnitude dispersion
in the strength of the bumps within a given sub-type. These dispersions
are tabulated for the LMC templates in \citet{2006Crowther}.

Given these relatively large intrinsic dispersions, and the possibility of
WR stars experiencing lesser amount of extinction with respect to the 
surrounding nebula, we consider only those cases that required templates of
two or more distinct WR subtypes as sure cases of multiple systems.
For these reasons WR4 and WR21 are considered to be individual stars, in spite
of having $f_{\rm WR}>1$ with $A_{\rm V}(nebula)$.
The template used for both these cases is that of WN2--4.
On the other hand, each of WR3 and WR6 required distinct templates (WNL and WCE), 
and hence are genuine multiple systems.
The multiplicative factors using $A_{\rm V}(Gal)$, the inferred WR subtype, and 
the number of WR stars at each of our 21 locations
are tabulated in Table~\ref{tab:comparison}.

There are 
only two compelling cases for the presence of multiple stars within our slits. These are
the same two cases (WR3 and WR6) for which we have inferred the
presence of multiple WR stars in Paper I, using the same LMC templates, but
using observed spectra without dereddening corrections.
We find 3 cases that show a weak RB, but with RB luminosity not exceeding 20\% of the
WCE template. The BB clearly suggests the presence of nitrogen features.
We classify these 3 stars as transitional type (WN/C), following
the definition of \citet{1989Conti}. 
Two of these transitional WRs (WR\,11, WR\,14) required WNL templates,
while WR\,5, required WNE template.
Accordingly, we name them WNL/CEs and WNE/CEs, respectively. 
Red bump strength of these stars is lower than that of  
the LMC transitional star BAT99-36, but are stronger than the red bump
strength sometimes seen in WN stars \cite[e.g. BAT99-26 in the LMC;][]{1999Breysacher}.

\begin{table}
\begin{center}
\caption{\label{tab:comparison} Classification of WR stars with LMC templates.}
\begin{tabular}{lcl}
\hline
ID     & $f_{\rm WR}\times$Template & WR subtype \\
(1)    & (2)                   & (3) \\
\hline
WR\,1  &   0.5 WC4             & 1 WCE \\
WR\,2  &   0.9 WC4             & 1 WCE \\
WR\,3  &   3 WN7--9+2 WC4      & 3 WNL+2 WCE \\
WR\,4  &   1.6 WN2--4          & 1 WNE  \\ 
WR\,5  &   1.5 WN2--4/0.1 WC4  & 1 WNE/CE \\
WR\,6  &   1 WN7--9+0.7 WC4    & 1 WNL+1 WCE \\
WR\,7  &   0.6 WN7--9          & 1 WNL \\
WR\,8  &   0.5 WN2--4          & 1 WNE \\
WR\,9  &   0.36 WC4            & 1 WCE \\
WR\,10 &   0.19 WC4            & 1 WCE \\
WR\,11 &   1 WN7--9/0.15 WC4   & 1 WNL/WCE \\
WR\,12 &   0.9 WN2--4          & 1 WNE \\
WR\,13 &   1.2 WN7--9          & 1 WNL \\
WR\,14 &   0.2 WN7--9/0.15 WC4 & 1 WNL/WCE \\
\hline
New 7\\
WR\,15 &   0.65 WN2--4         & 1 WNE \\
WR\,16 &   0.1 WC4             & 1 WCE \\
WR\,17 &   0.35 WC4            & 1 WCE \\
WR\,18 &   0.2 WC4             & 1 WCE \\
WR\,19 &   0.8 WN2--4          & 1 WNE \\
WR\,20 &   1 WN7--9            & 1 WNL \\
WR\,21 &   0.8 WN2--4          & 1 WNE \\
\hline
\end{tabular}\\
(1) WR identification number;
(2) Multiplicative factors for each template (observed spectra dereddened using $A_{\rm V}=0.20$~mag);
(3) Classification using LMC templates;
The number corresponds to the number of WRs of the given sub-type. 
\end{center}
\end{table}
\begin{table}\  
\begin{center}
\caption{\label{tab:elines} Emission lines contributing to the WR
features. 
}
\begin{tabular}{ccccccccc}
\hline
{ID} &     & Ion     &{$\lambda_0$}& \multicolumn{2}{c}{WN} & \multicolumn{2}{c}{WC} &  WO \\
(1)  & (2) & (3)     & (4)         & \multicolumn{2}{c}{(5)}   &\multicolumn{2}{c}{(6)}  & (7)\\
\hline
{BB} &     &         &             & L  & E     & L  & E  &   \\
1 & WR     & \heii\  & 4686        & a  & a     & o  & s  & o \\
2 & WR     & \niii\  & a)4634      & o  & n     & n  & n  & n \\
   &        & \niii\  & b)4641     & o  & n     & n  & n  & n \\
3 & WR     & \nv\    & a)4603      & n  & w     & n  & n  & n \\
   &        & \nv\    & b)4620     & n  & w     & n  & n  & n \\
4 & WR     & \ciii\  & 4647/66     & n  & n     & o  & o  & o \\
5 & WR     & \civ\   & 4658        & n  & n     & o  & o  & a \\
6 & Neb    & \hei\   & 4713        &    &       &    &    &   \\
7 & Neb    & \heii\  & 4686        &    &       &    &    &   \\
{RB}&      &         &             &    &       &    &    &   \\
8 & WR     & \ciii\  & 5696        & n  & n     & a  & n  & n \\
9 & WR     & \civ\   & 5801/12     & w  & w     & o  & a  & a \\
10& Neb    & \nii\   & 5755        &    &       &    &    &   \\
11& Neb    & \hei\   & 5876        &    &       &    &    &   \\
VB  &      &         &             &    &       &    &    &   \\
12& WR     & \ovi\   & 3811/34     & n  & n     & n  & n  & a \\
\hline
\end{tabular}\\
\end{center}
(1) Identification number of the Gaussian components in blue (BB), red (RB)
and a 'violet' (VB: bluer) bumps.
(2) Nature of the contributing emission line: WR (broad) or nebular
(narrow);
(3) The emitting ion;
(4) Rest wavelength in \AA;
(5-7) WR subtype where the corresponding emission line appears: always seen (a), 
often seen (o), weak (w) or never seen (n), as for the
information available at \citet{1968Smith, 1996Smith, 1998Crowther}.
Late and Early types for WN and WC are indicated by the column header
letters L and E, respectively.
\\
\end{table}

In summary, 19 of our spectra suggest the presence of individual stars
within our slit, with 3 WNLs (WR\,7, WR\,13 and WR\,20), 6 WNEs (WR\,4, WR\,8, WR\,12, WR\,15,
WR\,19 and WR\,21), 7 WCEs (WR\,1, WR\,2, WR\,9, WR\,10, WR\,16, WR\,17 and WR\,18), and 3
WN/Cs (WR\,5, WR\,11 and WR\,14). The remaining 2 cases contain
multiple WR stars (WR3 and WR6). An inspection at the {\it HST} images
suggests that the separation between these multiple stars should be
less than the PSF of the ACS camera, which is 1.8~pc at the distance
of M\,81.
In our analysis in Paper I, we had reported 4 transitional stars. We here note 
that two of those (WR\,5, and WR\,14) retain this classification, and two others
(WR\,9, and WR\,10) are reclassified as WCEs, with our new analysis, 
where we have taken into account reddening of the observed spectra.
Additionally, WR\,11, which was classified as WNL, is now re-classified as WN/C.

\section{ANALYSIS AND DISCUSSION}

WR features, the BB and RB, are usually blends of several ionic transitions.
The lines constituting the BB are resolved in some of our WN types.
When a WC component is present the lines are usually broader than 20\,\AA\ 
and the presence of the \ciii\ line in the BB often gives it a broad, unresolved appearance 
spanning wavelengths between $\sim$4600 to 4700\,\AA. On the other hand,
the RB is always broad and consists of \ciii\ and \civ\ lines. 
In addition, the adjoining nebular lines (\heineb\ in the BB; 
\niineb\ and \civheir\ in the RB) often blend into the bumps.
In Table~\ref{tab:elines} we list all the lines, both from the WR atmosphere 
and the surrounding nebula, that contribute to the bumps.

The individual lines contributing to the bumps can be recovered by using 
multiple-component Gaussian fitting. The decomposition would not only
allow us to check whether the recovered lines are consistent with its
classification made through LMC template fitting, but also allow us to
determine the line parameters such as flux and width.
Hence, we carried out multi-component Gaussian analysis of bumps.

\begin{figure*}
\begin{centering}
\includegraphics[width=1.0\linewidth]{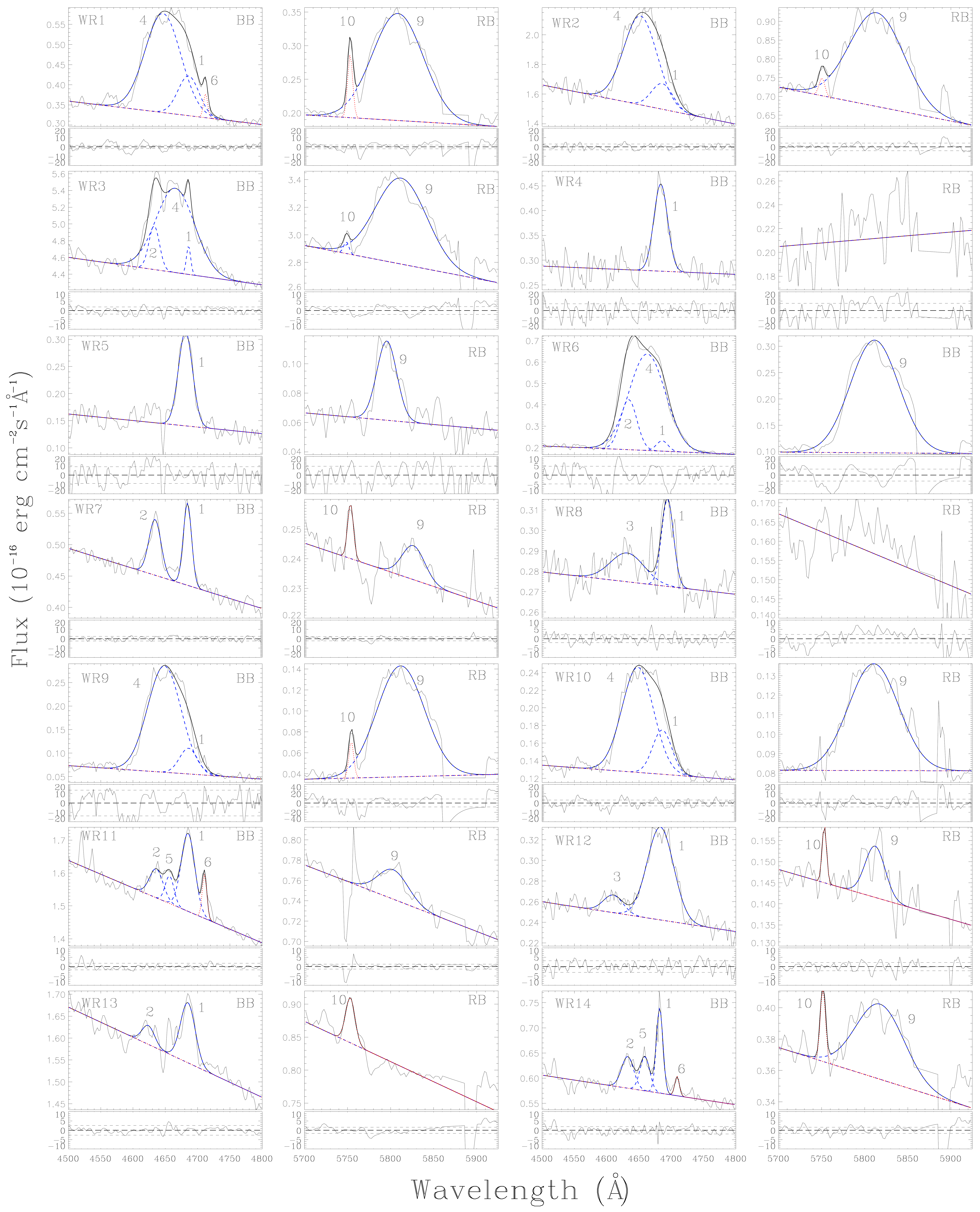}~
\par\end{centering}
  \caption{Multi-component Gaussian fits to the blue ({\it left}) and red ({\it right}) bumps: 
broad lines (dashed blue); nebular (dotted red when present); the sum (black). The fitted continuum
is shown by the dashed straight line. The spectrum is identified by its WR number, 
and the fitted components are identified by the numbers 1--12 (see column 2 of Table~4).
* indicates \ciii/\civ. Residuals are shown at the bottom of each panel.
}
\end{figure*}

\begin{figure*}
\begin{centering}
\includegraphics[width=1.0\linewidth]{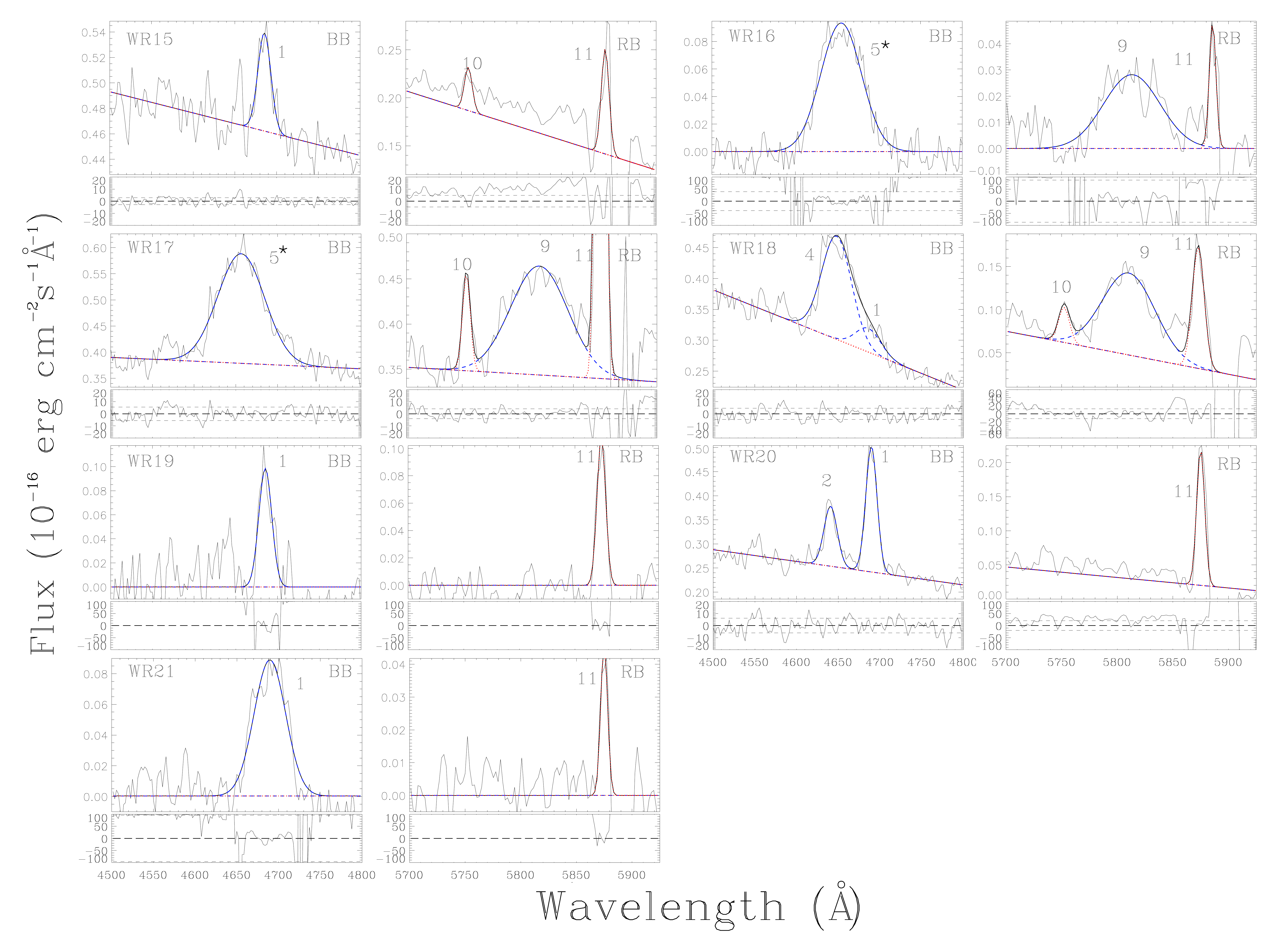}~
\par\end{centering}
Fig.~5 continued.
\end{figure*}

\subsection{Bump luminosities of M\,81 WRs}
We carried out the multiple-component Gaussian analysis of all our WR
spectra with a tailor-made code developed for that purpose which uses the {\sc idl}
routine {\sc lmfit}\footnote{The {\sc lmfit} function (lmfit.pro) performs a
non-linear least squares fit to a function with an arbitrary
number of parameters. It uses the Levenberg-Marquardt algorithm,
incorporated in the routine {\it mrqmin} of Numerical Recipes in C:
The Art of Scientific Computing (Second Edition), Cambridge University Press.}. 
The fittings were carried out on the spectra dereddened by $A_{\rm V}(Gal)$. 
We used the lines listed in Table~\ref{tab:elines} to define Gaussians
centered at the rest wavelengths ($\lambda_{0}$) of individual transitions.
The $\lambda_{0}$ values of the transitions were taken from
the {\sc nist} atomic spectra 
database\footnote{http://physics.nist.gov/PhysRefData/ASD/}.
For each bump, a continuum is defined by linearly interpolating on the
red and blue sides of the bumps, avoiding any lines in these parts of
the spectrum. Two of the 3 parameters, the peak intensity $I_0$ and
the line width $\sigma$ that defines each Gaussian, were left
free. Any line with a full width at half maximum (FWHM)$>9$~\AA\,
(that is $\sigma>4$~\AA), which is the typical value for the
instrumental profile, is defined as a broad line associated with a WR
star. The third parameter, $\lambda_0$, is fixed to one of the
values listed in Table~\ref{tab:elines}. 

The fitting program is executed interactively, where the visually
noticeable nebular lines were fitted first, using a Gaussian of 
FWHM$\sim$9\,\AA, appropriate to the instrumental profile. 
For the BB, the \heii\ broad
component is fitted next. Residuals are examined for a peak near any
of the other listed lines. If present, a second broad line is fitted
and residuals were re-examined. While fitting this second line, the
$\sigma$ and $I_0$ of the first line were left free. The process of
examining the residuals and adding a new line is continued until the
residual is indistinguishable from the noise in the spectrum. 
In none of the cases, the RB required more than one broad component.
In the iterative process, any faint nebular lines, if needed, were fitted. 

The results of Gaussian fittings of WR spectra are shown in Figure~5.
Structure of the BB is found to be complex, with a few objects fitted with 
a single Gaussian, while the majority required at least two Gaussians. 
For example, WR\,5 shows the most simple case: a single Gaussian was needed;
WR\,7 shows the case where two Gaussians were needed which are 
easily identified as independent components, even by eye; 
WR3 and WR6 shows a BB without any noticeable peak,
which is best fitted with 3 Gaussians, corresponding to \niii, \ciii\ and \heii\ lines.
Given the limitations in obtaining unique values of the fitted Gaussian
when more than 2 components are required, we fixed the FWHM of the \ciii\ line in the BB
to the corresponding value for the \civ\ line in their RB. Other cases where
more than 2 lines were required, individual lines are partially resolved and hence
their FWHM parameters of the fit are left free.

Multi-Gaussian fittings to the BB and RB gave the luminosity, FWHM and Equivalent 
Width (EW) of all the fitted Gaussians for each WR source of the sample.
The parameters of the broad \heiiwr\ line in the BB and the \civwrr\ line
in the RB are tabulated in Table~\ref{tab:param1}. Parameters for 
other broad lines in the BB are given in Table~\ref{tab:param2}.
Broad \heiiwr\ line is present in the BB in all but two cases.
The two sources without the \heiiwr\ line are WR\,16 and WR\,17, where
the BB is made up of the \ciii/\civ\ line. RB is detected in these two
sources and hence they are clearly WC stars. 

\begin{table*}
\begin{center}
\caption{\label{tab:param1}Parameters of \heii\ and \civ\ line from multi-component Gaussian fitting ($A_{\rm V}=0.20$~mag extinction correction)}
\begin{tabular}{l|crrrr|clrrrrr}
\hline
      & \multicolumn{5}{c}{\heii}   |& \multicolumn{5}{c}{\civ} |& \multicolumn{2}{c}{$L_{bumps}$}    \\
      \hline
ID   
&$\lambda_{0}$&$L$& FWHM & EW & S/N&$\lambda_{0}$&$L_{n(RB)}$& FWHM & EW & S/N & $L_{BB}$ |&$L_{RB}$ \\
(1)    & (2)  & (3) & (4) & (5) & (6)|& (7)  & (8) & (9) &(10)  &(11)&(12)   & (13)\\
\hline
WR\,1  & 4686 & 6.4 &38.5 &12.5 &60.6 & 5808 &0.56 &69.9 & 62.8 &48.5& 33.3  &18.7 \\
WR\,2  & 4686 &11.3 &39.0 & 4.8 &32.3 & 5814 &0.43 &74.6 & 29.6 &24.5& 72.2  &31.3 \\
WR\,3  & 4686 & 4.4 & 7.6 & 0.6 &51.7 & 5812 &0.58 &70.0 & 16.7 &48.9&126.9  &73.3 \\
WR\,4  & 4684 & 7.4 &25.0 &16.9 &14.0 & --   &0.06*& --  & --   &13.1&  7.4  & 0.5*\\
WR\,5  & 4681 & 7.5 &25.6 &33.9 &11.2 & 5796 &0.33 &27.2 & 25.5 &10.6&  7.5  & 2.5 \\
WR\,6  & 4686 & 1.7 &20.9 & 5.9 &21.3 & 5811 &0.37 &66.2 &154.1 &15.5& 63.5  &23.8 \\
WR\,7  & 4684 & 3.3 &15.1 & 4.8 &39.7 & 5827 &0.09 &31.0 &  1.6 &43.8&  6.6  & 0.65 \\
WR\,8  & 4694 & 1.6 &21.8 & 3.6 &46.7 & --   &0.04*& --  & --   &44.0&  3.0  & 0.1 \\
WR\,9  & 4686 & 3.0 &32.7 &34.1 & 7.3 & 5812 &0.44 &65.0 &197.1 &11.9& 26.1  &11.5 \\
WR\,10 & 4686 & 2.9 &34.9 &15.0 &32.1 & 5810 &0.44 &68.5 & 49.1 &24.1& 14.5  & 6.3 \\
WR\,11 & 4685 & 9.9 &24.8 & 4.2 &56.9 & 5803 &0.13 &41.1 &  1.7 &74.8& 15.8  & 2.0 \\
WR\,12 & 4683 & 7.3 &48.1 &19.0 &32.3 & 5812 &0.07 &27.3 &  2.5 &45.8&  8.3  & 0.6 \\
WR\,13 & 4685 & 6.2 &26.7 & 2.5 &38.7 & --   &0.06*& --  & --   &60.2&  7.8  & 0.4*\\
WR\,14 & 4682 & 3.4 &12.0 & 3.8 &51.6 & 5818 &0.69 &66.2 &  9.5 &60.5&  7.8  & 5.3 \\
WR\,15 & 4685 & 2.4 &18.3 & 3.3 &34.6 & --   &0.14*& --  & --   &20.8&  2.4  & 0.3*\\
WR\,16 & --   & --  & --  & --  & 2.5 & 5813 &0.31 &59.5 & --   & 1.2&  9.0  & 2.8 \\
WR\,17 & --   & --  & --  & --  &17.8 & 5819 &0.53 &61.0 & 22.8 &24.3& 23.3  &12.3 \\
WR\,18 & 4686 & 2.3 &35.6 & 5.2 &21.4 & 5810 &0.63 &55.8 &120.2 & 8.2& 14.1  & 8.9 \\
WR\,19 & 4685 & 3.0 &18.2 & --  & 1.0 & --   &0.07*& --  & --   & 0.9&  3.0  & 0.2*\\
WR\,20 & 4690 & 6.9 &15.9 &18.2 &16.4 & --   &0.03*& --  & --   & 5.0& 10.9  & 0.3*\\
WR\,21 & 4690 & 6.6 &44.6 & --  & 1.0 & --   &0.02*& --  & --   & 0.6&  6.6  & 0.1*\\
\hline
\end{tabular}\\
\end{center}
(1) WR identification;
(2) Observed center wavelength of \heiiwr\ [\AA];
(3) Luminosity [$1.0\times10^{35}$~\ergs];
(4) Full Width at Half Maximum (FWHM) [\AA];
(5) Equivalent Width (EW) [\AA];
(6) Continuum Signal to Noise Ratio (S/N) at 4750-4830\,\AA;
(7) Center of \civwrr\ feature;
(8) \civwrr\ luminosity normalized to $L_{BB}$;
(9) FWHM [\AA];
(10) EW [\AA];
(11) S/N at 5600-5730\,\AA.\\
(12) BB luminosity ($L_{BB}$) [$1.0\times10^{35}$~\ergs];
(13) RB luminosity ($L_{RB}$) [$1.0\times10^{35}$~\ergs]. * indicates 3$\sigma$ upper limits.
\end{table*}

\begin{table*}
\begin{center}
\caption{\label{tab:param2}Parameters of the 2$^{\rm nd}$ and 3$^{\rm rd}$ Gaussian components ($A_{\rm V}=0.20$~mag extinction correction).} 
\begin{tabular}{rrrrrrrrr}
\hline
      & \multicolumn{4}{c}{\niii-\nv} |& \multicolumn{4}{c}{\ciii-\civ}\\
      \hline
ID&$\lambda_{0}$&$L_n$& FWHM & EW &$\lambda_{0}$& $L_n$ & FWHM & EW \\
(1)    & (2)  & (3)  & (4)  & (5) & (6)  & (7)  & (8)  & (9) \\
\hline
WR\,1  & --   & --   & --   & --  & 4646 & 0.81 & 65.1 & 51.6\\
WR\,2  & --   & --   & --   & --  & 4652 & 0.84 & 60.5 & 25.3\\
WR\,3  & 4633 & 0.03 & 19.3 & 0.6 & 4665 & 0.93 & 70.0 & 17.0\\
WR\,4  & --   & --   & --   & --  & --   & --   & --   & --  \\
WR\,5  & --   & --   & --   & --  & --   & --   & --   & --  \\
WR\,6  & 4634 & 0.19 & 30.0 &39.3 & 4663 & 0.79 & 66.2 & 170.3\\
WR\,7  & 4634 & 0.50 & 22.3 & 4.7 & --   & --   & --   & --  \\
WR\,8  & 4632 & 0.48 & 61.5 & 3.4 & --   & --   & --   & --  \\
WR\,9  & --   & --   & --   & --  & 4649 & 0.89 & 60.9 &247.4\\
WR\,10 & --   & --   & --   & --  & 4648 & 0.80 & 57.6 & 57.7\\
WR\,11 & 4636 & 0.23 & 24.6 & 1.5 & 4657 & 0.14 & 15.8 &  0.9\\
WR\,12 & 4610 & 0.12 & 35.6 & 2.6 & --   & --   & --   & --  \\
WR\,13 & 4624 & 0.21 & 22.9 & 0.7 & --   & --   & --   & --  \\
WR\,14 & 4632 & 0.30 & 22.1 & 2.5 & 4659 & 0.26 & 17.8 & 2.2 \\
WR\,15 & --   & --   & --   & --  & --   & --   & --   & --  \\
WR\,16 & --   & --   & --   & --  & 4655 & 1.00 & 57.3 & --  \\
WR\,17 & --   & --   & --   & --  & 4656 & 1.00 & 66.1 & 39.0\\
WR\,18 & --   & --   & --   & --  & 4648 & 0.84 & 42.1 & 25.0\\
WR\,19 & --   & --   & --   & --  & --   & --   & --   & --  \\
WR\,20 & 4641 & 0.36 & 18.7 & 9.8 & --   & --   & --   & --  \\
WR\,21 & --   & --   & --   & --  & --   & --   & --   & --  \\
\hline
\end{tabular}\\
(1) WR ID;
(2) Observed center wavelength of \niiiwr\ or \nvwr\ [\AA];
(3) $L_\mathrm{NIII-NV}$ normalized to $L_{BB}$;
(4) FWHM [\AA];
(5) EW [\AA];
(6) Observed center wavelength of \ciii/\civ\ [\AA];
(7) $L_\mathrm{CIII-CIV}$ normalized to $L_{BB}$;
(8) FWHM [\AA];
(9) EW [\AA].
\end{center}
\end{table*}

\begin{figure*}
\begin{centering}
\includegraphics[width=0.75\linewidth]{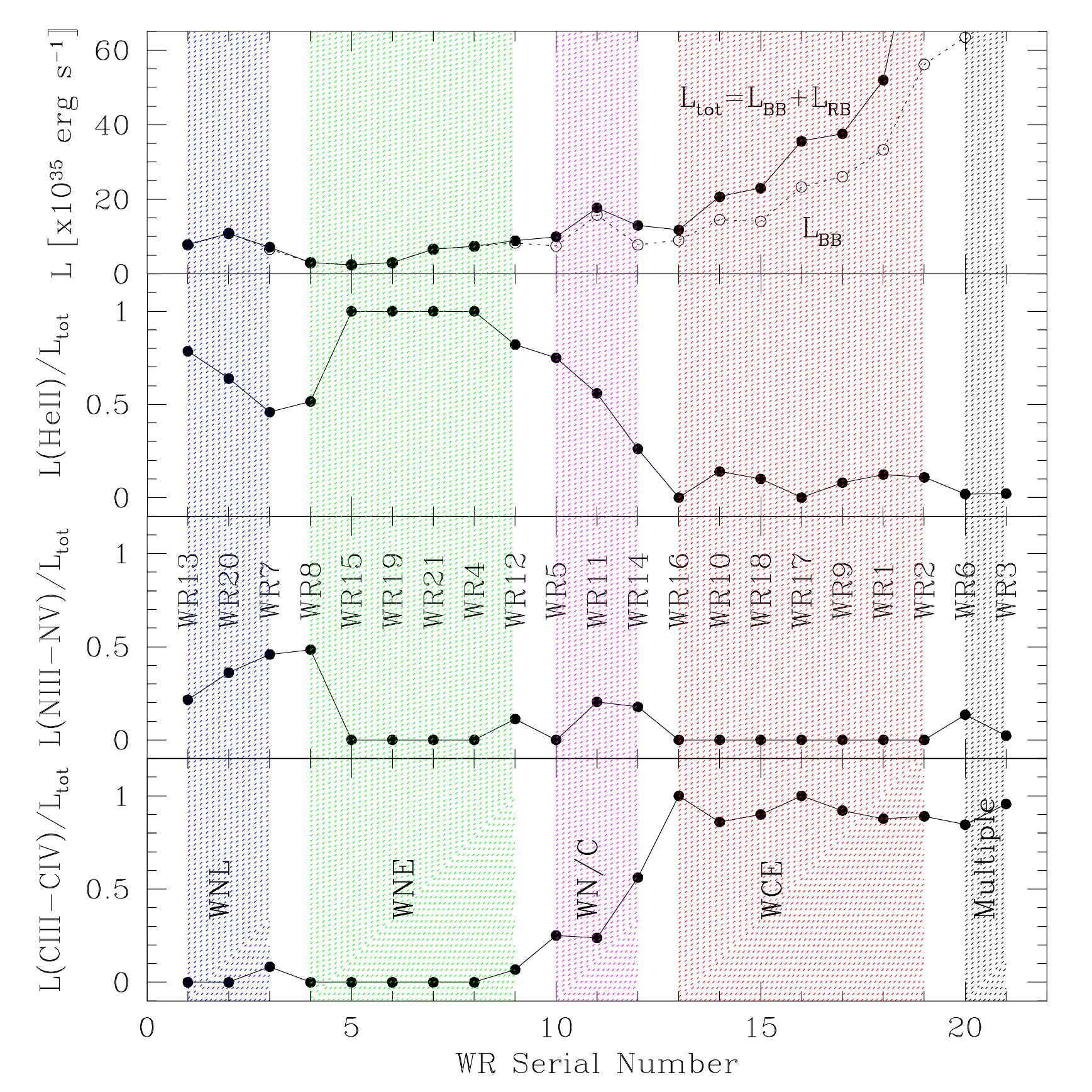}
\par\end{centering}
\caption{
Gaussian-decomposed luminosities of He, N, and C ions as a function of WR sub-type
for the 21 M\,81 WRs, which are grouped into WNL (blue), WNE (green), transitional (magenta),
WCE (red) subtypes. Locations with multiple WRs are shown by the black hatched area to the right.
Individual WRs are identified in the second panel from the bottom. Top panel shows
the summed luminosity ($L_{\rm tot}$) of the two bumps (solid circles and lines) and the blue bump luminosity (dotted line
joining empty circles). The next 3 panels show luminosities of He, N and C lines, all normalized to $L_{\rm tot}$.
}
\end{figure*}

The recovered lines from Gaussian fittings for all sources are in agreement 
with the lines expected for their classification (see Table~\ref{tab:elines}).
For example all WNLs show a nitrogen line in the BB (in addition to the \heii\ line) 
and WCs show one of the C lines (\ciiiwrb\ or \civwrb) in the BB, in addition 
to the \civwrr\ line in the RB.
The recovered N line in all WNLs is the \niiiwr\ line. 
All WNEs are not expected to show N lines, and when it is present
it is usually the \nvwr\ line.
The only apparent exception is the WNE source WR8. 
We detect both the N and C lines in sources classified as transitional.
In two of these transitional objects (WR\,11 and WR\,14), a C line 
is recovered in addition to the \civwrr\ line in the RB. 
This further justifies the transitional nature of this component.

The resultant Gaussian fit to the \heiiwr\ line agrees with the rest wavelength
of this line within $\sim$2\,\AA\ for all cases except WR\,8, WR\,20 and WR\,21.
Given that each spectrum was Doppler-corrected using the mean velocities of the nebular 
lines in the same spectrum, this agreement suggests that the relative 
radial velocity between the WR star and the surrounding ionized gas 
is less than 125\,\kms. Dispersion in the central wavelength of the other
components is larger, which is expected as several transitions at slightly
different wavelengths contribute to these components. 

The \heiiwr\ line is redshifted by 8\,\AA\ in WR\,8 and 4\,\AA\ in WR\,20 and WR\,21
(see $\lambda_{0}$ in Table~5),
suggesting that these objects are moving radially at velocities of
500 and 250\,\kms, respectively, 
with respect to the surrounding ionized gas.
The large Doppler velocity of WR\,8, resolves the apparent inconsistency
of the inferred N line for this WNE star. The recovered $\lambda_0=4632$
is closer to the \niiiwr\ rather than the \nvwr\ line.
Taking into account its Doppler shift and the large FWHM of the Gaussian fit,
the recovered \niii\ line is consistent with the \nvwr\ line, which is 
characteristic of WNE stars.

The trends of the relative strengths of the He, N and C 
lines for different spectroscopic WR sub-types are summarized in Figure~6.
We sorted the WR sub-types following the sequence suggested by the 
single-star evolutionary models. Thus WNLs are to the left (shown by the 
blue hatches), followed by WNEs (green), transitional (magenta) and WCEs (red).
Multiple stars are positioned at the extreme right.
We choose the sum of the luminosities of blue and red bumps for normalizing
the luminosities of individual transitions. In the upper panel
we show this total luminosity ($L_{\rm tot}$) as well as the BB 
luminosity for all our WR sources.
WNL and WNE stars have only the BB and hence the two luminosities are the same.
Within each sub-class, our WR stars are arranged in the increasing order of
$L_{\rm tot}$. A clear trend of increasing $L_{\rm tot}$ as a function
of evolutionary stage can be seen. All spectra showing a RB have at least
$L_\mathrm{tot}=4\times10^{35}$\,\ergs. 
The remaining three panels show the luminosities of
\heii, N and C lines, all normalized to $L_{\rm tot}$.
From WNL to WNE, the N fraction decreases, and in WCEs,
the \heii\ is weak or absent. These trends are exactly as expected in the 
massive star evolutionary scenario of WRs. 
The transitional stars have properties intermediate to WN and WC stars.

\begin{figure*}
\begin{centering}
\includegraphics[width=0.8\linewidth]{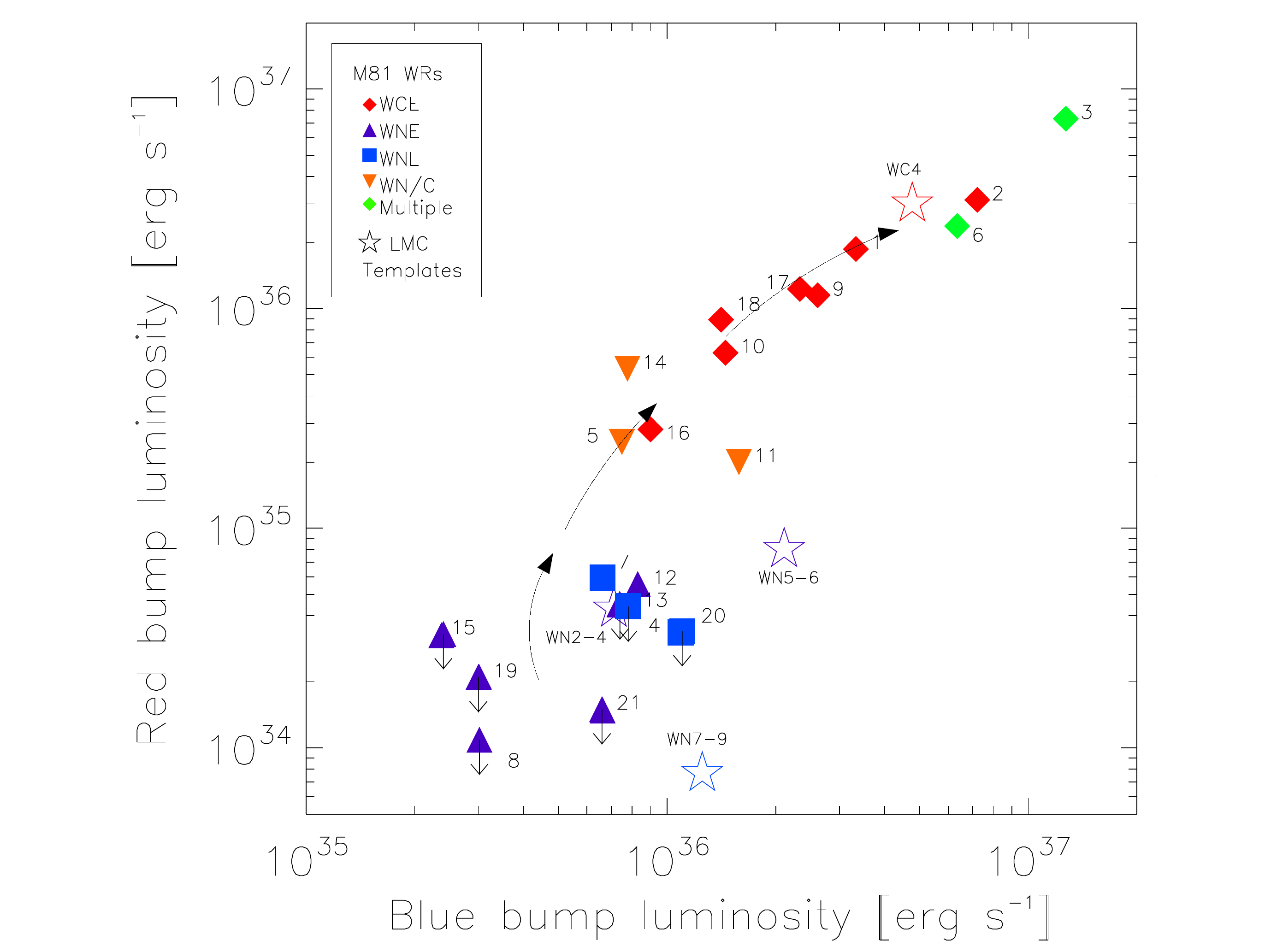}~
\par\end{centering}
  \caption{RB vs BB luminosity diagram for the 21 WR stars in M\,81.
A 3$\sigma$ upper limit is indicated where no RB was detected.
Their spectral identification is indicated, as well as their
classification using observational templates of individual WR stars
in the LMC: WNL (blue box), WNE (purple triangle), WCE (red
diamond), transitional WN/C (orange inverted triangle) and multiple stars (green diamond).
LMC observational templates from different WR-types are shown
overplotted with filled and empty stars.
Arrows indicate the expected trajectory of single star evolutionary models
in this diagram.
}
\end{figure*}

In Paper~I, we showed that a diagram involving RB luminosity vs BB
luminosity is able to separate different WR sub-types. This diagram
has the potential to help classify extra-galactic WR stars, where
it is often difficult to identify all key ionic transitions that are 
commonly used for classification of LMC WRs.
Furthermore, it is a common practice to use a typical luminosity of 
$1.6\times10^{36}$~\ergs\ for the BB to estimate the number of
WRs in WR galaxies and other distant starburst systems \citep{Schaerer1998}.
This typical luminosity is based on the WRs in the LMC and the Milky Way and
needs to be verified in other extragalactic systems. We use the M\,81 WR data
to discuss the luminosities of the blue and red bumps and their
dependence on the WR sub-type.

The multi-Gaussian fitting has allowed us to decompose the bumps
into their individual ionic transitions. The method also provided us
luminosities in these individual ionic transitions, without the
contamination of nebular lines. The BB could contain one or more of
the following broad emission lines: \heiiwr, \niiiwr, \nvwr, \ciiiwrb\ and \civwrb.
So we summed the recovered luminosities (see Table~\ref{tab:param1} and Table~\ref{tab:param2})
of these lines to obtain the total BB luminosity.
For the case of the RB, we only included the luminosity of \civwrr. 
In seven cases (WR\,4, WR\,8, WR\,13, WR\,15, WR\,19,
WR\,20, WR\,21), where no RB was detected, a 3$\sigma_{\rm RB}$ upper
limit is used as the luminosity of the RB, where $\sigma_{\rm RB}$ 
was calculated assuming that the RB occupies $\sim$40~pixels ($\sim$80\,\AA;
$\sigma_{\rm RB}=\sqrt{40}\sigma$, where $\sigma$ is the continuum rms/pixel
close to the RB).

The summed luminosities of blue and red bumps are given in the last 2 
columns of Table~\ref{tab:param1} and are plotted in Figure~7.
The 7 WRs with only an upper limit for their RB luminosity 
are shown in the figure by a vertical arrow pointing downwards.
For comparison we measured the bump luminosities from the LMC templates, 
which are also shown in Fig.~7.
This RB vs BB luminosity diagram shows that the N-rich WR stars are located in
the lower-left luminosity range, whereas the C-rich WR stars are located towards
the upper-right region with higher luminosities.
Transitional objects, those classified as WN/C, lie at intermediate values. 
Finally, multiple systems exhibit the highest luminosities in both the RB and BB.
For all the
LMC WNL templates, the BB luminosities are clustered
around the often used value of $1.6\times10^{36}$~\ergs. On the other hand,
the two M\,81 WNLs have a value which is around half of that. 
The BB is systematically fainter for the M\,81 WNEs, and lies between
the values for WNEs of LMC templates.
For WCE stars, there is a large dispersion in the RB and BB luminosities
in M\,81 as well as the two templates used.
The observed luminosities for these stars in M\,81 are in the same range as 
that for the templates. 

Single star evolutionary models follow the scenario proposed by
\citet{1976Conti}, wherein WR stars are evolutionary products of O-stars.
As an O-star starts losing mass following the exhaustion of hydrogen (H) 
in its core, it starts revealing the products of CNO H-burning in its atmosphere. 
As a consequence the surface abundance of He, N, C and O start increasing,
at the cost of H \citep[][]{2005Meynet}. A post main-sequence O-star is 
considered to be a WR star when its surface abundance of H drops to 
$\sim$50\%. The first phase of a WR is the WNL. 
When the He abundance reaches $\sim$78\%, it is considered to be WNE
and when C abundance is greater than $\sim$20\% it is considered a WC star. 
Thus, the evolutionary sequence followed by WRs is from WNL to WNE to WCs.
We show this trajectory in Figure~7 by arrowed curves. 
The transitional WN/C stars lie in this trajectory intermediate to WNE and WC stars.

In single star evolutionary models without rotation, the surface N 
abundance drops to zero at the onset of the WC phase, and C abundance is
zero in the WN phase. In this scheme, a star
is not expected to simultaneously show N and C lines, and
hence these models cannot explain the transitional WN/C stars.
\citet{2005Meynet} found that stars rotating at initial speeds of $\sim$300\,km\,s$^{-1}$
were able to bring out the He-burning products from its core before removing 
completely the CNO-enriched envelope of stars with masses between $\sim$30--60~\msol. 
Hence, these stars evolve through the transitional WN/C phase for a non-negligible 
duration, during which time they show the broad lines of both N and C 
simultaneously. The location of our WN/C stars along the evolutionary
path between the WNE and WC phase is consistent with this scenario.

The relative locations of different sub-types follow the same trend in
LMC and M\,81. As discussed above these trends are consistent with
the sequence of occurrence of different sub-types in current single
star evolutionary models.
M\,81 is a 
star-forming galaxy, with a few hundreds of WRs expected in it 
and hence future searches would increase the sample size.
Nevertheless, we deem it relevant to carry out a deeper analysis of our present
dataset that would enable us to make quantitative comparisons with the WRs
in other galaxies.
The characteristics of our transitional WN/C stars resemble
those of the WN subtypes in the BB, and those of the WC subtypes in the RB.
In addition, the RB fluxes of our WN/C are lower than those of the WCE stars.
Therefore, a multiple WN + WC system in which the BB would dominate the flux
and shape of a WC is discarded.

A WR detection in an extragalactic spectrum may contain contribution
of more than one star within the observed slit, either because they
are in a binary or they are part of a cluster. If the two stars are of
the same type (WN or WC), the combined spectrum would be a simple
multiplication of the spectrum of that type.
However, if the two stars are of different types, the
combined spectrum will contain both the N and C lines. Spectra of such
multiple systems can also mimic the spectra of rarely seen
transitional WN/C type. However, a detailed analysis of the spectra
can distinguish between these possibilities.
The spectra of three of our objects (WR\,5, WR\,11 and WR\,14) 
are classified as transitional WR stars: these objects clearly
show a residual indicating the need for a nitrogen line, when using
only a carbon-rich template and vise versa. The requirement for
a WCE in these stars are guided by the presence of RB in them. 
The strength of carbon line
in the RB in these 3 stars is much weaker as compared to that of
a typical WCE
and hence are unlikely to be multiple systems.
Multi-Gaussian fitting suggests the presence of a C line in
the BBs of WR\,11 and WR\,14. This reinforces the need for a WC component.
The observed BB of WR\,11 and WR\,14, could only be fitted by having
both WN and WC features. Specifically, without a WN feature there is a
systematic residual at the blue edge of the BB.
Although the BB of WR\,5 looks like WN sub-type,
the observed RB could only be
produced by having also a WC feature.

\section{Summary and concluding remarks}

In this work, we reported the detection of WR features at 7 new locations in
the nearby spiral galaxy M\,81. These detections were the result of
spectroscopic observations of the central stars of {\it HST}-detected ionized
bubbles using the GTC/OSIRIS spectrograph at the 10.4-m GTC. Using the
same instrumental set-up,
we had earlier detected 14 locations with WR features in this galaxy.
We carried out a detailed analysis of all the 21 WR spectra using
the well-established LMC templates
of WRs of different spectroscopic sub-types
with the intention of classifying and quantifying the WR responsible
for the observed bump strength.

In general,
the bump strengths do not require more than one WR star
in 19 cases of which 3 are WNLs (WR\,7, WR\,13 and WR\,20),
6 WNEs (WR\,4, WR\,8, WR\,12, WR\,15, WR\,19 and WR\,20),
7 WCEs (WR\,1, WR\,2, WR\,9, WR\,10, WR\,16, WR\,17 and WR\,18),
and 3 WN/Cs (WR\,5, WR\,11 and WR\,14). The remaining 2 cases contain
multiple WR stars (WR\,3 and WR\,6). Examination of the {\it HST} images
suggests that the separation between these multiple stars is
less than the PSF of the ACS camera, which is 1.8~pc at the distance
of M\,81. None of the detections correspond to WCL and WO types.
One of the detected WRs (WR\,8) is found to have high (500\,\kms)
radial velocity with respect to the surrounding ionized gas.

We analysed the observed bumps using multiple-component Gaussian fitting
in order to recover the ionic transitions responsible for the bumps.
In all cases, the recovered ions are consistent with those expected for the
inferred sub-type using the templates. All WRs of a given sub-type occupy a
distinct zone in the RB vs BB luminosity diagram, with our transitional objects
having bump luminosities intermediate between that of WNs and WCs.
The BB luminosity of M\,81 WNs is at least a factor of 2 less than the
often used typical value of $1.6\times10^{36}$~\ergs\ based on the 
LMC WNs. No such difference is found for WC stars.
We suggest that RB vs BB diagram is a straightforward tool of
spectral classification of WR stars, especially for
extragalactic sources (single or in clusters) where detailed spectra
from each star cannot be obtained.

The detection of as much as 19 individual WR stars in M\,81, which is at
a distance of 3.6~Mpc, opens up a new environment for testing
the massive star evolutionary models.

\section*{Acknowledgements}
The authors thank the referee Paul Crowther for his enlightening comments
and suggestions improving this work.
To Antonio Cabrera and the rest of the GTC staff for
their help in carrying out the observations presented in this work and
also the support during data reductions. This work is based on
observations made with the NASA/ESA Hubble Space Telescope, obtained
from the data archive at the Space Telescope Science Institute. STScI
is operated by the Association of Universities for Research in
Astronomy, Inc. under NASA contract NAS 5-26555.
YDM thanks CONACyT for the research grant CB-A1-S-25070 and
DRG for the research grant CB-A1-S-22784.
VMAGG thanks to CONACyT, INAOE and UNACH, for the different supports to carry
out this work.
JAT and VMAGG are funded by UNAM DGAPA PAPIIT project IA100318.


\end{document}